\documentclass[final]{aipproc}

\def\selectedlayoutstyle{6x9} 
\layoutstyle\selectedlayoutstyle

\SetInternalRegister\hbadness{8000} 

\newcommand{\nn}{\nonumber}
\newcommand{\bmp}{\mbox{\bf p}}
\newcommand{\bmpp}{\mbox{\bf p}'}
\newcommand{\bmk}{\mbox{\bf k}}
\newcommand{\sSS}{\mbox{\footnotesize 1S}}

\def\O#1#2{\mbox{ $O$}_{{\bf #1},#2}}
\def\Od#1#2{\mbox{ $O$}^\dagger_{{\bf #1},#2}}
\def\bsigma{\bf\sigma}
\def\psip#1{\psi_{\bf #1}}
\def\chip#1{\chi_{\bf #1}}

\newcommand\doingARLO[2][]{%
  \ifx\mmref\undefined #1\else #2\fi
}

\begin{document}

\title[]{Threshold Top Quark Production
\footnote{Plenary talk at the 9th International Symposium on Heavy Flavor 
Physics, Sept. 2001}
}


\author{Iain W. Stewart}{
address={\begin{center}
Department of Physics, University of California at San Diego,\\[2pt] 
9500 Gilman Drive, La Jolla, CA 92093-0319, USA\end{center}},
email={iain@schwinger.ucsd.edu},
thanks={}
}
\copyrightyear{2001}

\begin{abstract}
Predictions for the total cross section for $e^+e^-\to t\bar t$ near
threshold are reviewed. The renormalization group improved results at NNLL order
have improved convergence and reduced scale dependence relative to fixed order
results at NNLO. Prospects for measurements of the top-quark mass, width, and
Yukawa coupling are discussed.
\end{abstract}

\date{\today}

\maketitle

\section{Introduction}

The production of top quark pairs is among the important projects of a future
linear collider. The large top quark width $\Gamma_t\sim 1.5$~GeV makes the
threshold cross section look quite different from that of charm or bottom
pairs. The large value of the width prohibits the production of toponium states,
and at the same time serves as an infrared cutoff from sensitivity to
non-perturbative effects. Thus, perturbative methods can be used to describe the
top-antitop dynamics to a very high degree of precision.  This makes the
threshold region an ideal place for extracting fundamental top quark parameters
such as the top mass, width, and Yukawa coupling (for a light higgs).

In the threshold region $\sqrt{s}\simeq 2m_t\pm 10\,{\rm GeV}$, the $t$ and
$\bar t$ move with non-relativistic velocities. Defining the energy $m_t {\rm
v}^2 = \sqrt{s}-2 m_t$ we see that this region of $s$ corresponds to velocities
$|{\rm v}|$ \raisebox{-3pt}{$\stackrel{<}{\sim}$} $\alpha_s$. In this region
an exact treatment of QCD Coulomb singularities $(\alpha_s/{\rm v})^k$ is
required, ruling out a pure $\alpha_s$ expansion. However, a combined expansion
in powers of ${\rm v}$ and $\alpha_s$ can be performed.  Schematically, for the
$e^+e^-\to t\bar t$ cross section, $\sigma_{t\bar t}(s)$, we want an expansion
of the form
\begin{eqnarray} \label{RNNLO}
 && R \, = \, \frac{\sigma_{t\bar t}}{\sigma_{\mu^+\mu^-}}
 \, = \, {\rm v}\,\sum\limits_k \left(\frac{\alpha_s}{\rm v}\right)^k \times
 \Big[\: 1\ +\ \{\alpha_s, {\rm v}\} \ +\  
 \{\alpha_s^2, \alpha_s {\rm v}, {\rm v}^2\}+ \ldots \,\Big] \,.\\[-2pt]
 && \hspace{5cm} \ \mbox{LO}\quad\ \mbox{NLO}\qquad\quad  \mbox{NNLO} \nn
\end{eqnarray}
The power counting for these corrections can be implemented in a simple and
systematic way using the effective theory framework of Non-Relativistic QCD
(NRQCD).

\begin{figure}
  {\includegraphics[width=0.5\textwidth]{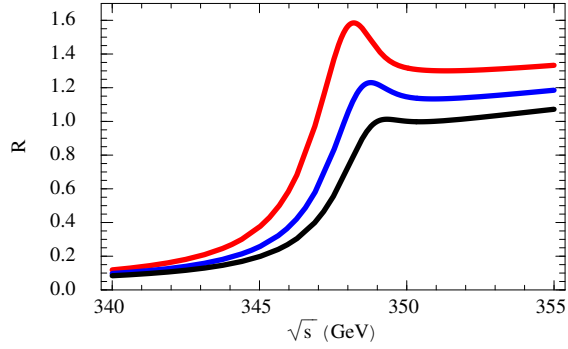}} 
  \caption{Leading order predictions for $R$. Results are shown
  for three renormalization scales of order the momentum transfer in the 
  Coulomb potential, $\mu=15\,{\rm GeV}$ (upper), $30\,{\rm GeV}$ (middle), 
  and $60\,{\rm GeV}$ (lower).}
  \label{fig_RLO}
\end{figure}

The leading order (LO) prediction for $R$ is shown in Fig.~\ref{fig_RLO}.
The large top width gives the threshold cross section a smooth line-shape, with
a single bump from the remnant of the 1S toponium state. The
characteristics of the cross section are sensitive to the top parameters. In
particular, the top mass determines the location of the rise/peak, the top width
determines the slope of the rise and shape of the peak, and the overall
normalization provides information on $\alpha_s(m_t)$ and the top-Yukawa
coupling for a light higgs. However, at LO the prediction for $R$ suffers from
considerable scale uncertainty as shown by the three curves in
Fig.~\ref{fig_RLO}.  This illustrates the importance of including higher order
terms. Several groups have computed the next-to-next-to-leading order (NNLO) QCD
corrections to the total cross section (see Ref.~\cite{Hoang3} for a summary and
comparison of these results). Using ``threshold'' top-quark mass parameters, an
accurate prediction for the location of the peak in the cross section was
obtained. It was concluded that infrared safe top masses could be determined
with a precision of $<200$~MeV.  However, surprisingly the NNLO corrections to
the cross section normalization were big, with large scale uncertainty. The
residual scale dependence was estimated to make the normalization of the cross
section uncertain to $\approx 20\%$~\cite{Hoang3}. These large corrections
seemed to jeopardize measurements of the top width, strong coupling, and Yukawa
coupling.

To understand the source of this large scale dependence it is useful to recall
that the dynamics of the top-antitop system are governed by vastly different
energy scales. These quarks have mass $m_t\sim 175\,{\rm GeV}$, typical momenta
${\bf p}\simeq m_t {\rm v}\sim 25\,{\rm GeV}$, and energies $E\simeq m_t{\rm v}^2
\sim 4\,{\rm GeV}$. The scattering amplitudes therefore involve logarithms
\begin{eqnarray}
  \ln\Big(\frac{\mu^2}{m_t^2}\Big)\,,\qquad 
  \ln\Big(\frac{\mu^2}{{\bf p}^2}\Big)\,,\qquad
   \ln\Big(\frac{\mu^2}{E^2}\Big)\,,
\end{eqnarray}
which are not all small for a single choice of $\mu$. In the expansion for
$R$ these large logarithms appear as logs of the velocity. This suggests that a
better expansion might involve a summation of these logarithms
\begin{eqnarray} \label{RNNLL}
 && R \, = \,{\rm v}\,\sum\limits_k \left(\frac{\alpha_s}{\rm v}\right)^k
 \sum\limits_j \big(\alpha_s\ln {\rm v} \big)^j \times
 \Big[\: 1\ +\ \{\alpha_s, {\rm v}\} \ +\  
 \{\alpha_s^2, \alpha_s {\rm v}, {\rm v}^2\}+ \ldots \,\Big] \,.\\[-2pt]
 && \hspace{5.3cm} \ \mbox{LL}\quad\ \mbox{NLL}\qquad\quad  \mbox{NNLL} \nn
\end{eqnarray}
The additional summation of logarithms can be performed using renormalization
group equations (RGE). This is similar to how logarithms of $m_W/m_b$ are summed
for the electroweak Hamiltonian. In our case the summation is complicated by the
presence of two low energy scales that are coupled by the equations of motion,
$E={\bf p}^2/(2m_t)$. This complication can be dealt with by using a
renormalization group with a subtraction velocity
$\nu$~\cite{LMR,amis,amis2,amis3,hms}. In this framework there are two
renormalization group parameters in the effective Lagrangian: $\mu_S$ for soft
gluons and $\mu_U$ for ultrasoft gluons, where $\mu_S=m_t\nu$ and
$\mu_U=m_t\nu^2$. Running from $\nu=1$ to $\nu\sim {\rm v}$ sums logarithms of
${\rm v}$ and minimizes both $\ln(\mu_S^2/{\bf p}^2)$ and $\ln(\mu_U^2/E^2)$
terms in the amplitudes. (The $\ln(\mu^2/m_t^2)$ terms are minimized by matching
QCD onto the effective theory near $\mu= m_t$.) In Refs.~\cite{hmst,hmst2} it
was shown that the expansion in Eq.~(\ref{RNNLL}) is better behaved than that in
Eq.~(\ref{RNNLO}). Furthermore, the normalization of the cross section at NNLL
has significantly smaller scale uncertainty ($\approx 3\%$) than the NNLO
result. This review focuses on these results; for more details and further
references see Ref.~\cite{hmst2}.

Below, the steps in the renormalization group improved calculation are briefly
described. This is followed by predictions for the cross section as well as a
description of the prospects for measurements of the top parameters in light of
the reduced theoretical uncertainties at NNLL order.

\section{Renormalization Group Improved Calculation}

The computation of the renormalization group improved cross section can be
divided into three parts:
\begin{enumerate}

\item Matching of QCD onto an effective theory for non-relativistic top quarks.
Determine the Wilson coefficients $C(\nu)$ at $\nu=1$ ($\mu=m_t$) to the desired
order in $\alpha_s(m_t)$.

\item Scaling $C(\nu)$ from $\nu=1$ to $\nu={\rm v_0}\simeq C_F\alpha_s$ by
calculating anomalous dimensions and using the renormalization group. This
scaling sums the terms of the form $[\alpha_s \ln({\rm v})]^j$ in
Eq.~(\ref{RNNLL}).

\item Computing the cross section using the Lagrangian and currents renormalized 
at the low scale $\nu={\rm v_0}$ .

\end{enumerate}
Each of these parts will be discussed in turn.

\subsection{1. Matching onto the Effective Theory}

For non-relativistic scattering the relevant momentum regions can be classified
by their typical energy and momenta ($k^0$,${\bf k}$). They include hard modes
with momenta $\sim(m,m)$, potential modes with momenta $\sim (m{\rm v^2},m{\rm
v})$, soft modes with momenta $\sim (m{\rm v},m{\rm v})$, and ultrasoft modes
with momenta $\sim (m{\rm v^2},m{\rm v^2})$.  Fluctuations involving
hard or offshell momenta are integrated out, while effective theory fields are
introduced for modes with nearly on-shell momenta. The degrees of freedom
therefore include potential top and anti-top quarks ($\psi_{\bf p}$ and
$\chi_{\bf p}$), soft gluons and light quarks ($A_q^\mu$ and $\varphi_q$), and
ultrasoft gluons and light quarks ($A^\mu$ and $\phi_{us}$). Soft energies and
momenta appear as labels on the fields while ultrasoft momenta are represented
by explicit coordinate dependence~\cite{LMR}. This enables us to distinguish
the size of momenta, for instance a derivative $\partial^\mu \psi_{\bf p}(x)\sim
m{\rm v^2} \psi_{\bf p}(x)$.

In this framework the action for non-relativistic top quarks has terms 
\begin{eqnarray} \label{Lke}
 {\mathcal L} &=& \sum_{\bf{p}}
   \psi_{\bf p}^\dagger(x)   \Biggl\{ i D^0 - {({\bf p}-i{\bf D})^2 \over 2 m} 
   +\ldots\Biggr\} \psi_{\bf p}(x) + (\psi\to \chi) \nn \\
 && - \sum_{\bf p,p'} F\!\left({\bf p} ,{\bf p^\prime}\right)\:
  \Big[ \psi_{\bf p^\prime}^\dagger(x)\:
  \psi_{\bf p}(x)\: \chi_{\bf -p^\prime}^\dagger(x) \:
  \chi_{\bf -p}(x) \Big] \nn\\
 && - 2\pi\,\alpha_s(m\nu) \sum_{{\bf p,p'},q,q'}
 \psi_{\bf p^\prime}^\dagger\: [A^\alpha_{q'},A^\beta_{q}] 
 U_{\alpha\beta}\:\psi_{\bf p} + (\psi\to \chi)\,,
\end{eqnarray}
where color and spin indices are suppressed. The covariant derivatives in the
first line involve only ultrasoft gluons. The function $F({\bf p},{\bf p'})$ in
the second line contributes to the potential between quarks and anti-quarks. For
our purposes
\begin{eqnarray} \label{F}
  F({\bf p},{\bf p'}) = \frac{1}{\bmk^2} {\cal{V}}_c(\nu)
    + \frac{\pi^2}{m |\bmk|}\, {\cal{V}}_k(\nu)
    + \frac{1}{m^2}{\cal V}_2(\nu) 
    +\frac{{\bf S}^2}{m^2} {\cal{V}}_s(\nu)
    +\frac{(\bmp^2+\bmpp^2)}{2 m^2 \bmk^2}\,{\cal{V}}_r(\nu)\,,
\end{eqnarray}
where $\bmk=\bmpp-\bmp$, ${\bf S}$ is the total spin operator, and all
coefficients are in the color singlet channel. The matching for these Wilson
coefficients is needed at two loops for ${\cal{V}}_c$, one loop for ${\cal
V}_k$, and tree level for ${\cal V}_{2,s,r}$~\cite{amis3,hms}.  Finally, the
third line in Eq.~(\ref{Lke}) is an example of the type of interaction that
occurs between potential quarks and soft gluons, with $U_{\alpha\beta}({\bf
p,p^\prime},q,q^\prime)$ a matching function of the label momenta. At NNLO
(and NNLL order) time ordered products of two of these soft interactions also
contribute terms to the potential giving
\begin{eqnarray} \label{Vsoft}
 V_{\rm soft}(\bmp,\bmpp) & = &
 -\,\frac{C_F\, \alpha_s^2(m\nu)}{\bmk^2}\,\bigg[\,
 -\beta_0\,\ln\Big(\frac{\bmk^2}{m^2\nu^2}\Big) + a_1 \,\bigg] \\ 
& & -\,\frac{C_F\, \alpha_s^3(m\nu)}{4\pi\,\bmk^2}\, \bigg[\,
 \beta_0^2\,\ln^2\Big(\frac{\bmk^2}{m^2\nu^2}\Big) - \Big(2\,\beta_0\,a_1 +
 \beta_1\Big)\,\ln\Big(\frac{\bmk^2}{m^2\nu^2}\Big) + a_2 \,\bigg] \,,\nn
\end{eqnarray}
where $\beta_i$ are coefficients of the QCD $\beta$-function and the constants
$a_i$ can be found in Ref.~\cite{Schroder}.  The complete potential is then
$V(\bmp,\bmpp) = F(\bmp,\bmpp) + V_{\rm soft}(\bmp,\bmpp)$.

We also need to take into account that the top quarks decay. We will assume that
the decay products are hard and can be integrated out. At lowest order this
induces the operators
\begin{eqnarray} \label{Gtop}
 {\cal L} = \sum_{\bf{p}} \psip{p}^\dagger \: \frac{i}{2} \Gamma_t\: \psip{p} 
  \ +\  \sum_{\bf{p}}   \chip{p}^\dagger\: \frac{i}{2} \Gamma_t\: \chip{p} \,.
\end{eqnarray}
In the Standard Model the dominant decay channel is $t\to bW^+$ and gives a
width of $\Gamma_t=1.43\,{\rm GeV}$ which we will use as our central
value. Counting $\Gamma_t\sim m_t {\rm v}^2$, Eq.~(\ref{Gtop}) gives a
consistent next-to-leading order treatment of electroweak effects for the total
cross section~\cite{NonFact1,NonFact2}.  Thus, we will not include electroweak
decay related effects to the same order as the QCD corrections. From partial
knowledge of these corrections~\cite{HT}, the missing terms are expected to be
at the few percent level.

Besides the effective Lagrangian we also need external currents to produce the
top-antitop pair. Since we wish to describe $e^+e^-\to \{\gamma^*,Z^*\} \to
t\bar t$ these currents are induced by both electromagnetic and weak
interactions.  The relevant vector current is ${\bf J}^{\rm v}_{\bf{p}}=
c_1(\nu) \O{p}{1} + c_2(\nu) \O{p}{2}$, where
\begin{eqnarray}\label{Ov}
 \O{p}{1} & = & {\psip{p}}^\dagger\, \bsigma(i\sigma_2)\, {\chip{-p}^*} \,, 
 \qquad
 \O{p}{2} =  \frac{1}{m^2}\, {\psip{p}}^\dagger\, 
    \bmp^2\bsigma (i\sigma_2)\, {\chip{-p}^*} \,, \nn
\end{eqnarray} 
and the relevant axial-vector current is ${\bf J}^a_{\bf{p}}= c_3(\nu) \O{p}{3}
$, where
\begin{eqnarray}\label{Oa}
 \O{p}{3} & = & \frac{-i}{2m}\, {\psip{p}}^\dagger\, 
      [\,\bsigma,\bsigma\cdot\bmp\,]\,(i\sigma_2)\,{\chip{-p}^*} \,. 
\end{eqnarray} 
The matching for these Wilson coefficients is needed at two loops for $c_1$ and
tree level for $c_{2,3}$. The two-loop matching for $c_1$ is scheme
dependent~\cite{SC1,SC2,SC3}, and in the $\overline{\rm MS}$ scheme with our
definition of the operators, can be found in Ref.~\cite{hmst2}.

\subsection{2. Renormalization group scaling}

To sum the $(\alpha_s\ln{\rm v})^j$ terms in $R$ we must determine the anomalous
dimensions for the Wilson coefficients ${\cal V}_{c,k,2,s,r}$ in Eq.~(\ref{F})
and the current coefficients $c_{1,2,3}$. The anomalous dimensions for ${\cal
V}_{c}$, ${\cal V}_{k}$, and ${\cal V}_{2,s,r}$ are required at three, two, and
one loop respectively.  These have been computed in
Refs.~\cite{amis,amis3,hms,Pineda0,PinedaRGE}, and due to mixing depend on the
one-loop running of the HQET terms up to $1/m^2$~\cite{Bauer}.  
These anomalous dimensions can contain terms like (with $b_i$ coefficients that
depend on color factors)
\begin{eqnarray}
 \nu {\partial \over \partial\nu} {\cal V} &=& b_1
  \alpha_s^2(m_t\nu) +b_2 \alpha_s(m_t\nu) \alpha_s(m_t\nu^2)
  + b_3 \alpha_s^2(m_t\nu) \ln\Big[ \frac{\alpha_s(m\nu^2)}
  {\alpha_s(m\nu)} \Big] \, \nn\\ 
 &&+ b_4 \, c_F^2(\nu)\:\alpha_s^2(m_t\nu)+\ldots \,.
\end{eqnarray}
For ${\cal V}_{c,k,2,r}$ both soft and usoft loops contribute, so their
anomalous dimensions depend on both $\alpha_s(m_t\nu)$ and
$\alpha_s(m_t\nu^2)$. The anomalous dimension for ${\cal V}_s$ comes only from
soft loops but also involves the mixing of the $\bar\psi\, {\bf\sigma}\cdot {\bf
B}\, \psi$ operator whose Wilson coefficient is $c_F(\nu)$.
\begin{figure}
  {\includegraphics[width=0.5\textwidth]{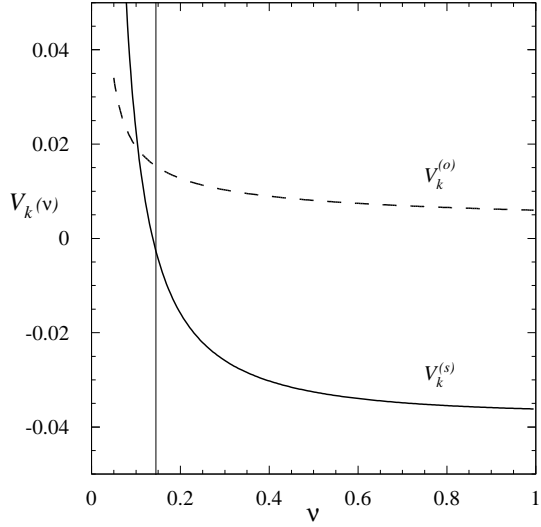}} \caption{Running of the
  color singlet (s) and octet (o) Wilson coefficients for the $\pi^2/(m|\bmk|)$
  potential from Ref.~\cite{amis3}. The solid vertical line marks the Coulombic 
  region where $\nu\simeq {\rm v_0}$.}  \label{pltvm1}
\end{figure}
Of the current coefficients, $c_3$ has no anomalous dimension, while $c_2(\nu)$
has contributions only from ultrasoft loops~\cite{hmst}.  More interesting is
the anomalous dimension for $c_1(\nu)$ which starts at two-loop order from
purely potential loops. At this order~\cite{LMR}
\begin{eqnarray} \label{adc1}
 \nu {\partial  \over \partial\nu} \ln[c_1(\nu)] 
 &=& -{{\cal V}_c(\nu)  \over 16\pi^2} \left( { {\cal V}_c (\nu) \over 4 }
  +{\cal V}_2 (\nu)+{\cal V}_r (\nu) + 2 {\cal V}_s(\nu)  \right) +
  { {\cal V}_{k}(\nu) \over 2} \,,
\end{eqnarray}
so the solution for $c_1$ depends on the solutions for the potential
coefficients. At three loops there are new contributions to the anomalous
dimension for $c_1(\nu)$ coming from mixed potential-ultrasoft and
potential-soft loops. These contribute at NNLL order but are currently unknown.
In Ref.~\cite{hmst2} these unknown terms were estimated to affect the cross
section at the 2\% level (this rough estimate was based on the size of known
terms and using dimensional analysis and parameter dependence to estimate the
size of the contributions which are unknown).  Of all the Wilson coefficients
the one which is the most responsible for the difference in the NNLO and NNLL
cross sections is ${\cal V}_k(\nu)$ which changes by an order of magnitude
between $\nu=1$ to $\nu=0.15$. The two-loop running of this coefficient is shown
in Fig.~\ref{pltvm1}.

\subsection{3. Cross section computation}

The total cross section for $e^+e^-\to t\bar t$ is given by
\begin{eqnarray}
  \sigma_{\rm tot}^{\gamma,Z}(s) = \frac{4\pi\alpha^2}{3s}
  \Big[\, F^v(s)\,R^v(s) +  F^a(s) R^a(s) \Big] \,,
\label{totalcross}
\end{eqnarray}
where $F^v$ and $F^a$ are trivial functions depending on the charge and weak
isospin of the fermions, and $\sin\theta_W$. $R^v$ and $R^a$ are determined by
\begin{eqnarray} \label{Rveft}
 R^v(s) & = & \frac{4\pi}{s}\,
 \mbox{Im}\Big[\,
 c_1^2(\nu)\,{\cal A}_1({\rm v},m,\nu) + 
 2\,c_1(\nu)\,c_2(\nu)\,{\cal A}_2({\rm v},m,\nu) \,\Big] \,,
\\[4mm] \label{Raeft}
 R^a(s) & = &  \frac{4\pi}{s}\,
 \mbox{Im}\Big[\,c_3^2(\nu)\,{\cal A}_3({\rm v},m,\nu)\,\Big] \,,\nn
\end{eqnarray}
where ${\cal A}_i$ are time-ordered products of effective theory currents [$\hat
q=(\sqrt{s}-2m_t,0)$]
\begin{eqnarray} \label{Ai}
 {\cal A}_1({\rm v},m,\nu) & = & i\,\sum\limits_{\bf p,p^\prime}
 \int\! d^4x\: e^{i \hat{q} \cdot x}\: \Big\langle\,0\,\Big|\, 
 T\, \O{p}{1}(x){\Od{p'}{1}}(0) \Big|\,0\,\Big\rangle \,, \nn
\\
 {\cal A}_2({\rm v},m,\nu) & = & \frac{i}{2}\,  \sum\limits_{\bf p,p^\prime}
 \int\! d^4x\: e^{i \hat{q}\cdot x}\: \Big\langle\,0\,\Big|\,
 T\,\Big[ \O{p}{1}(x){\Od{p'}{2}}(0)+\O{p}{2}(x){\Od{p'}{1}}(0)
 \Big] \Big|\,0\,\Big\rangle \,, \nn
\\
 {\cal A}_3({\rm v},m,\nu) & = & i\,  \sum\limits_{\bf p,p^\prime}
 \int\! d^4x\: e^{i \hat{q}\cdot x}\: \Big\langle\,0\,\Big|\, 
 T\, \O{p}{3}(x){\Od{p'}{3}}(0)\Big|\,0\,\Big\rangle \,.
\end{eqnarray}
These time-ordered products can be evaluated in terms of non-relativistic Greens
functions to give
\begin{eqnarray} \label{Gi}
  {\cal A}_1({\rm v},m,\nu) & = & 18\,\bigg[\, G^c({\rm v},m,\nu)
  + \left({\cal{V}}_2(\nu)+2{\cal{V}}_s(\nu)\right)\, 
  \delta G^\delta({\rm v},m,\nu) \nonumber \\
 & & \hspace{0.1cm} + \,{\cal{V}}_r (\nu)\,\delta G^r({\rm v},m,\nu) 
  + \,{\cal{V}}_k(\nu)\, \delta G^k({\rm v},m,\nu) 
  + \,\delta G^{\rm kin}({\rm v},m,\nu) \,\bigg] \,,\nn\\
  {\cal A}_2({\rm v},m,\nu) & = & {{\rm v}^2}\,{\cal A}_1({\rm v},m,\nu) \,,
  \qquad {\cal A}_3({\rm v},m,\nu)  =  {12}\, G^1({\rm v},m,\nu)/{m^2}\,,
\end{eqnarray}
Here $G^c$ are evaluated numerically with the $1/{\bf k}^2$ term in $F({\bf
p},{\bf p}')$ and $V_{\rm soft}({\bf p},{\bf p}')$~\cite{JKT}. In
Ref.~\cite{hmst2} we analytically evaluated $\delta G^{\delta,r,k,kin}$ with a
single insertion of the corresponding potentials or ${\bf p}^4$ kinetic energy
correction. The P-wave Greens function $G^1$ was also evaluated in closed
form. The analytic calculations enabled all ultraviolet subdivergences to be
subtracted in $\overline{\rm MS}$ which is necessary to be consistent with the
scheme dependence of the Wilson coefficients.  In Eq.~(\ref{Gi}) the velocity
${\rm v}=[\sqrt{s}-2m_t+i\Gamma_t]^{1/2}/m_t^{1/2}$ and $m=m_t$ is the pole
mass.  The Greens functions depend on the subtraction velocity through
$\ln(\nu^2/{\rm v}^2)$ and these logarithms are not large when the Greens
functions are evaluated at the low scale $\nu\simeq {\rm v}_0$. At this scale
all large logarithms have been resummed in the potential and current Wilson
coefficients. Typically, ${\rm v_0}\simeq 0.15-0.2$, but to numerically test the
remaining scale dependence we use the larger range ${\rm v}_0=0.1-0.4$.

\section{Results}

Soon after the NNLO results were derived it was realized that the inherent
uncertainty in the top-quark pole mass due to infrared renormalons causes
problems for predictions for the peak in the cross section. Therefore, for
precision predictions the pole mass is not a suitable mass parameter. The
$\overline{\rm MS}$ top-mass provides an infrared safe alternative, however it
complicates the non-relativistic power counting. Essentially, it shifts $E$ by
an amount $\sim m_t\alpha_s$ which is much larger than the original size of the
energy $E\sim m_t\alpha_s^2$. Both of these problems can be addressed by
switching to a ``threshold mass'', defined as an infrared safe mass parameter
which differs from $m_t^{\rm pole}$ by $\sim m_t\alpha_s^2$. Several possible
threshold masses were suggested, including the PS mass~\cite{PS}, kinetic
mass~\cite{KM}, and 1S mass~\cite{a1S,HT}. In Ref.~\cite{Hoang3} it was
concluded that threshold masses could be determined with a precision of
$<200$~MeV from the total cross section. Converting the result to an
$\overline{\rm MS}$ top-mass would then lead to a similar precision for this
parameter. In this section predictions will be given using the 1S mass
parameter. For a detailed description of how the NNLL pole mass expressions 
are converted to the 1S mass in a manner consistent with the power counting see
Ref.~\cite{hmst2}.

\begin{figure} \label{fig_compare}
  {\includegraphics[width=0.5\textwidth]{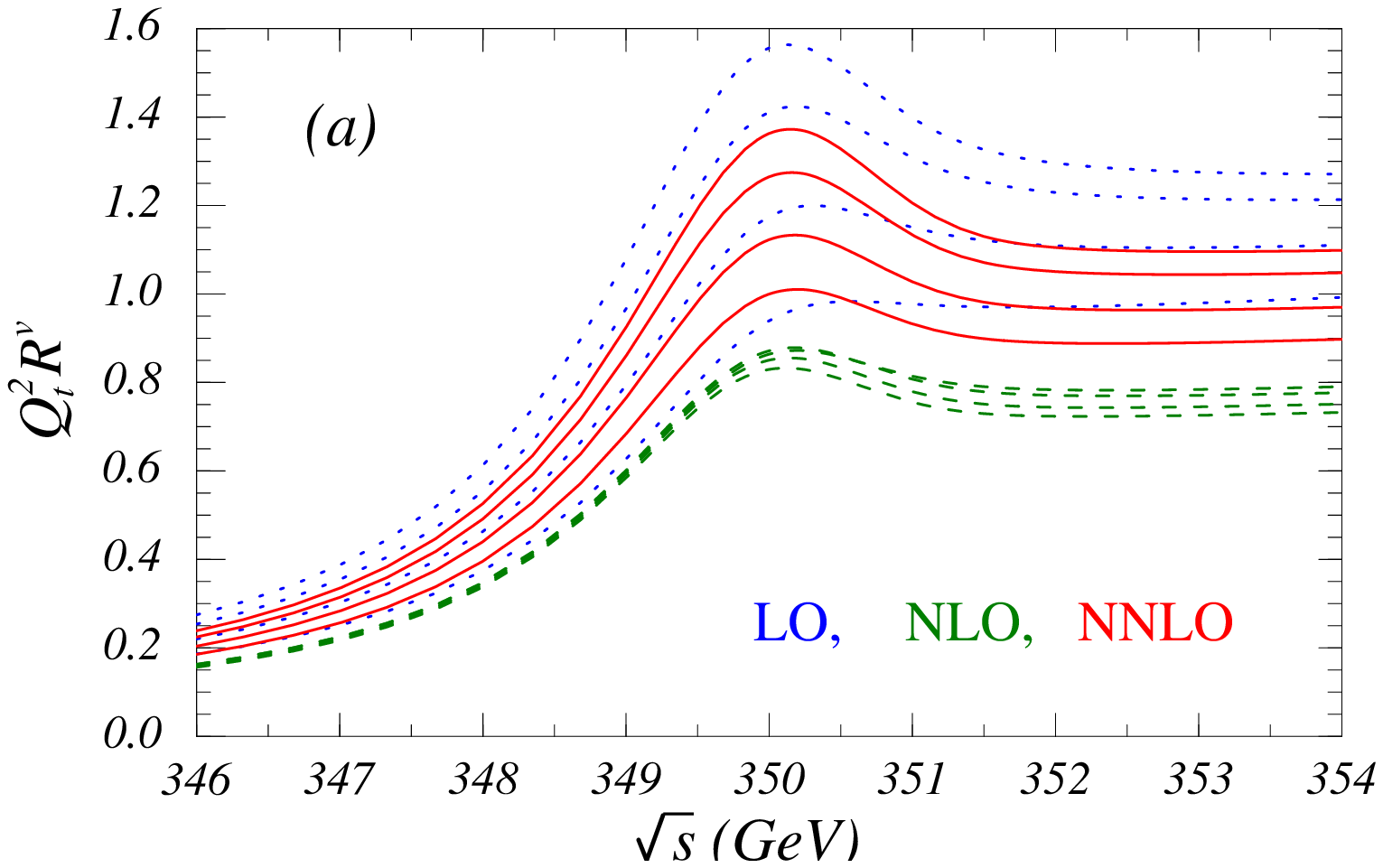}\hspace{0.2cm}
  \includegraphics[width=0.5\textwidth]{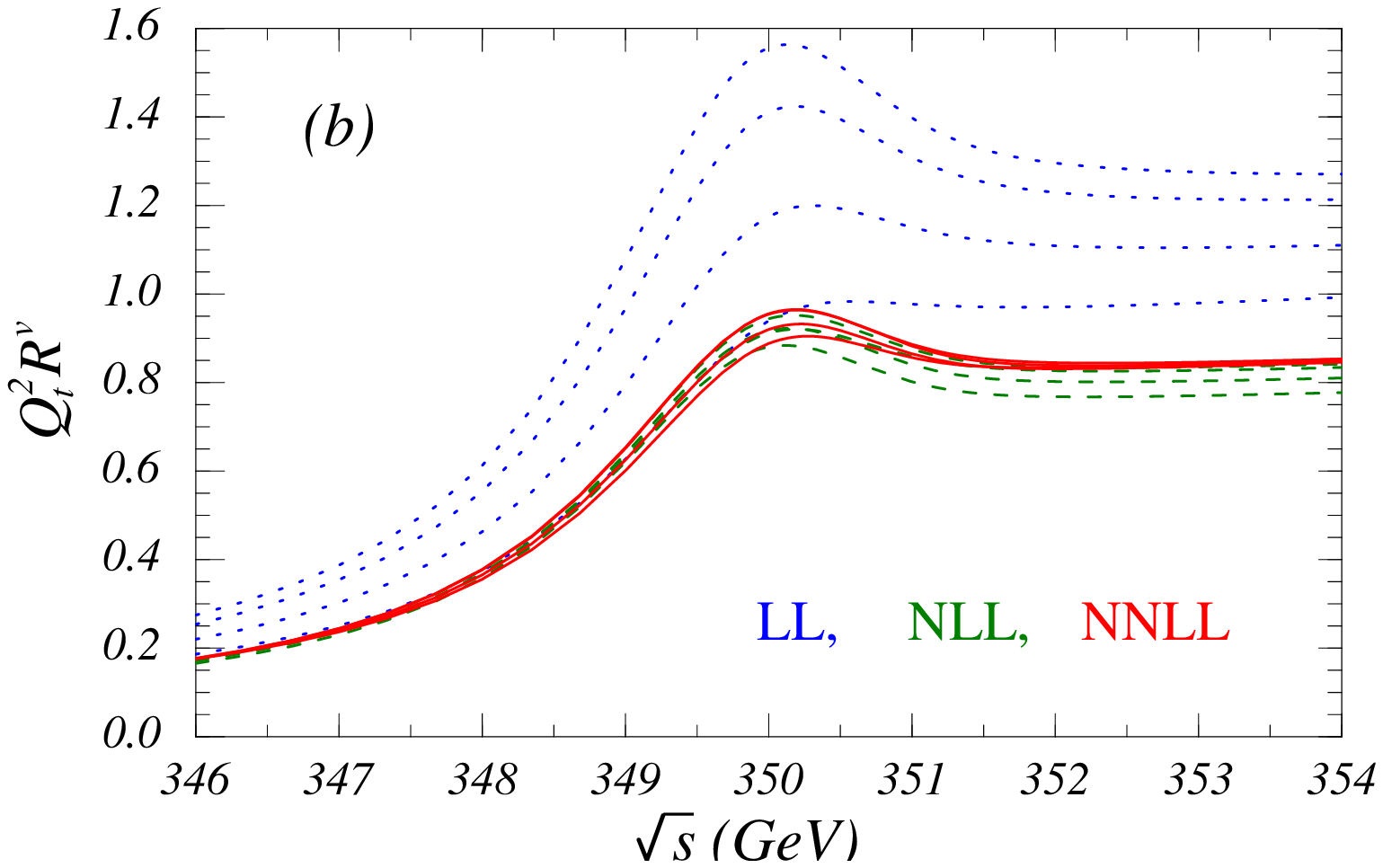}}
  \caption{Comparison of $Q_t^2 R^v$ with fixed $M_t^{\sSS}$ mass for
the fixed order and resummed expansions. The dotted, dashed, and solid
curves in a) are LO, NLO, and NNLO, and in b) are LL, NLL, and NNLL order.  For
each order four curves are plotted for $\nu=0.1$, $0.125$, $0.2$,
and $0.4$.}
\end{figure}

We begin by comparing results in the fixed order and renormalization group
improved expansions. We concentrate on $R^v$ since $R^a$ gives only a small
contribution to the cross section, and is essentially identical in the two
approaches.  The $R^v$ results are shown in Fig.~\ref{fig_compare} and use
$M_t^{\sSS}=175$~GeV, $\alpha_s(m_Z)=0.118$ and $\Gamma_t=1.43$~GeV. At each
order in the expansions four curves are shown which correspond to $\nu=0.1$,
$0.125$, $0.2$, and $0.4$. It is clearly visible that the NNLL results in
Fig.~\ref{fig_compare}(b) have much smaller scale dependence than the NNLO
results in Fig.~\ref{fig_compare}(a). It should be noted that our NNLO results
shown in Fig.~\ref{fig_compare}(a) agree quantitatively with those presented in
Ref.~\cite{Hoang3}. The uncertainty in these results stems to a large extent
from the uncertainty in the choice of the renormalization scales in the NNLO
contributions. Essentially what the anomalous dimensions and renormalization
group do is remove this uncertainty. Also, more than half of the improved
convergence of the NNLL result is due to the reduced size of ${\cal
V}_k(\nu={\rm v}_0)$ compared to ${\cal V}_k(1)$.

From the remaining scale uncertainty and the size of some higher order QCD
corrections, the uncertainty in the NNLL cross section was conservatively
estimated to be $\pm 3\%$~\cite{hmst2}.  This level of precision should enable
extractions of various top parameters from the cross section with fairly good
precision.
\begin{figure}
  \leavevmode {\includegraphics[width=0.5\textwidth]{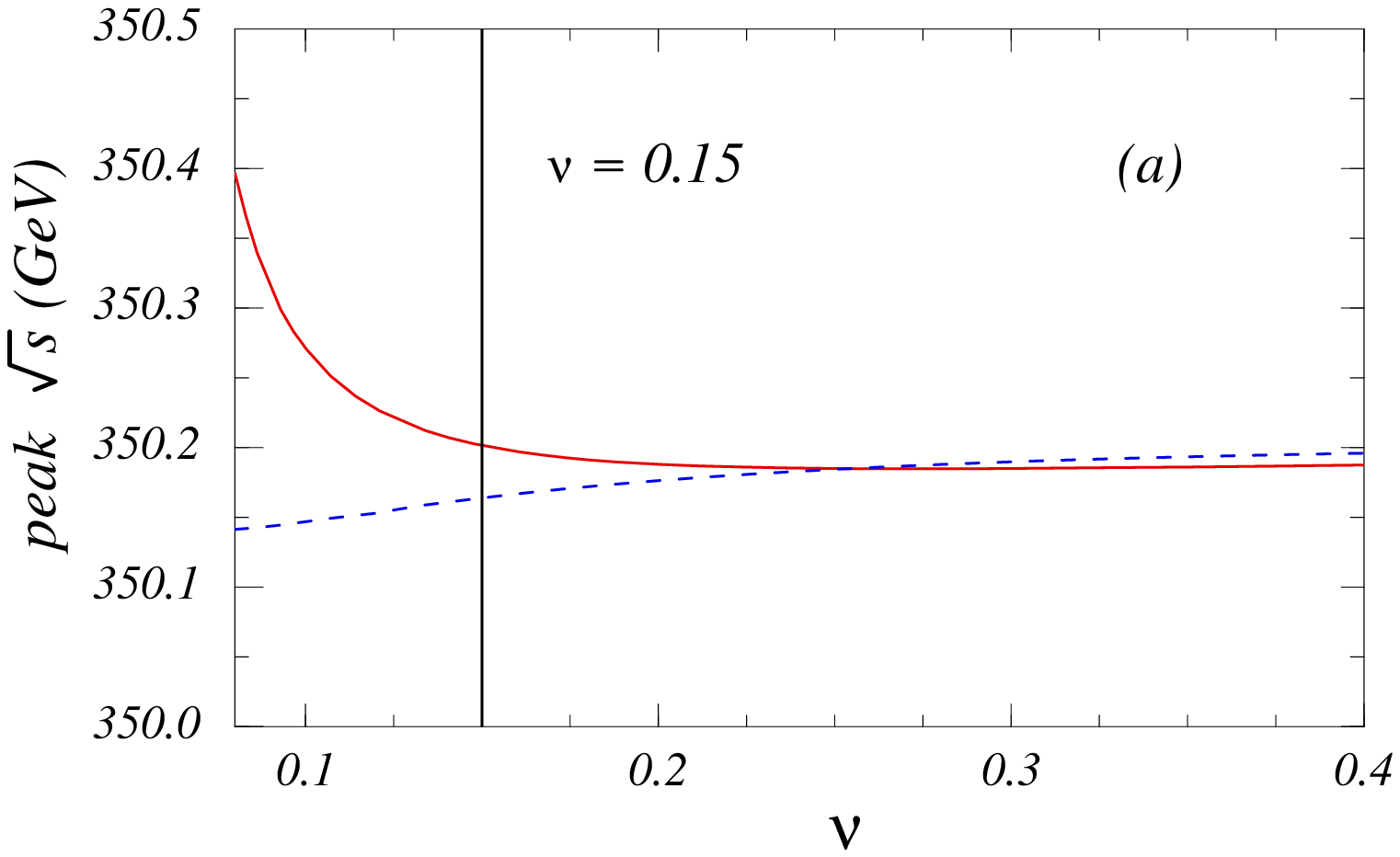}\hspace{0.2cm}
  \includegraphics[width=0.5\textwidth]{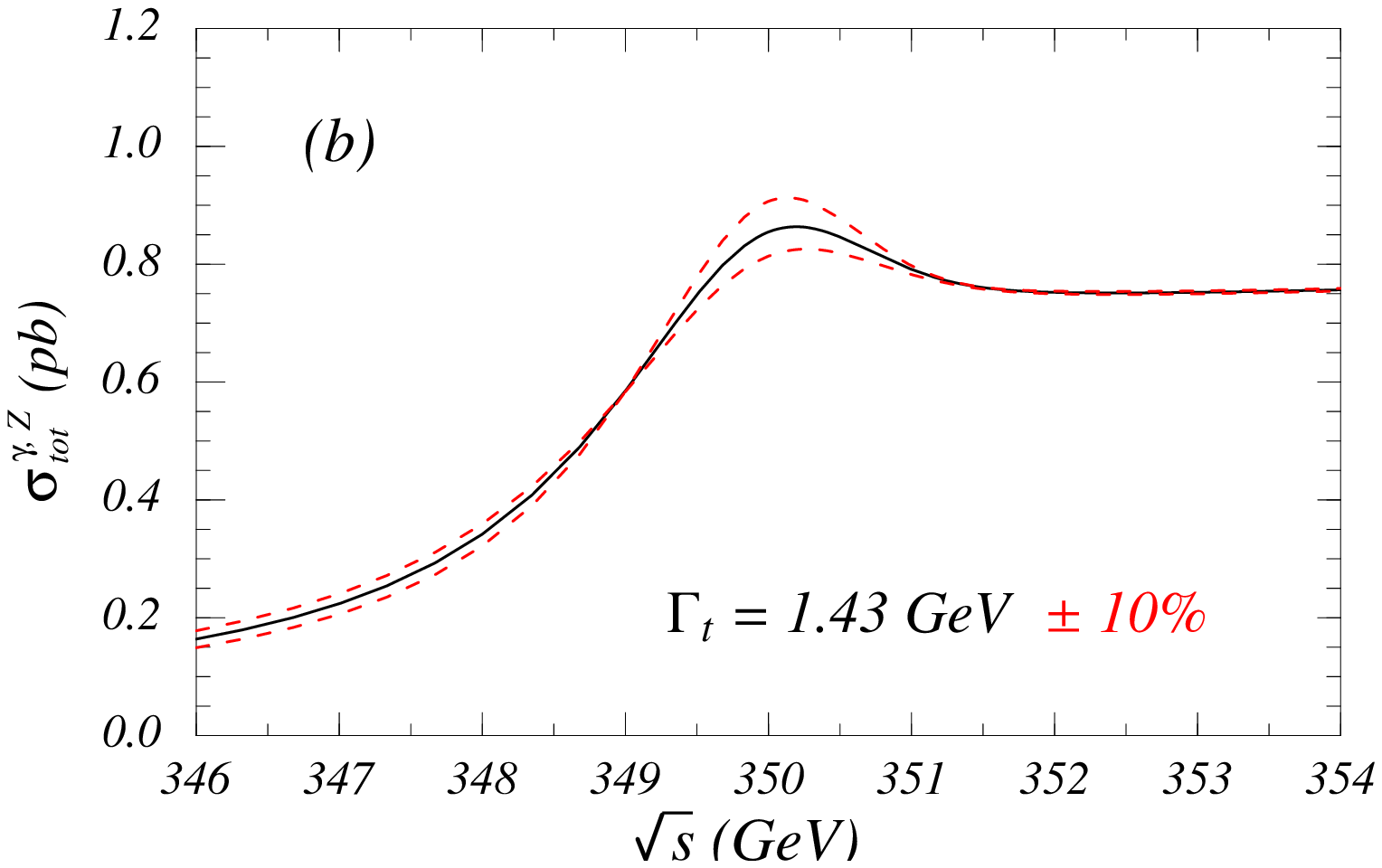}} \caption{(a) Position of the
  peak in the cross-section versus $\nu$ at NNLO (dashed) and NNLL (solid). The
  vertical line at $\nu=0.15$ is a physically motivated endpoint for the
  running. (b) Variation of the NNLL cross section for a $\pm 10\%$ change in 
  the value of the top quark width.}
\label{fig_params1}
\end{figure}
In Fig.~\ref{fig_params1} we show the scale dependence of the peak position for
the NNLO (dashed) and NNLL (solid) predictions. The NNLL prediction is slightly
less scale dependent than the NNLO prediction until we get to small $\nu$. For
$\nu<0.15$ the larger scale dependence at NNLL is explained by the fact that
these predictions depend on the coupling $\alpha_s(m_t\nu^2)$, while the NNLO
predictions do not. Also shown in Fig.~\ref{fig_params1} are the NNLL
predictions for the total cross section varying the width by $\pm 10\%$. The
size of the variations indicate that a measurement with better than 10\%
precision is definitely feasible.
\begin{figure}
  {\includegraphics[width=0.5\textwidth]{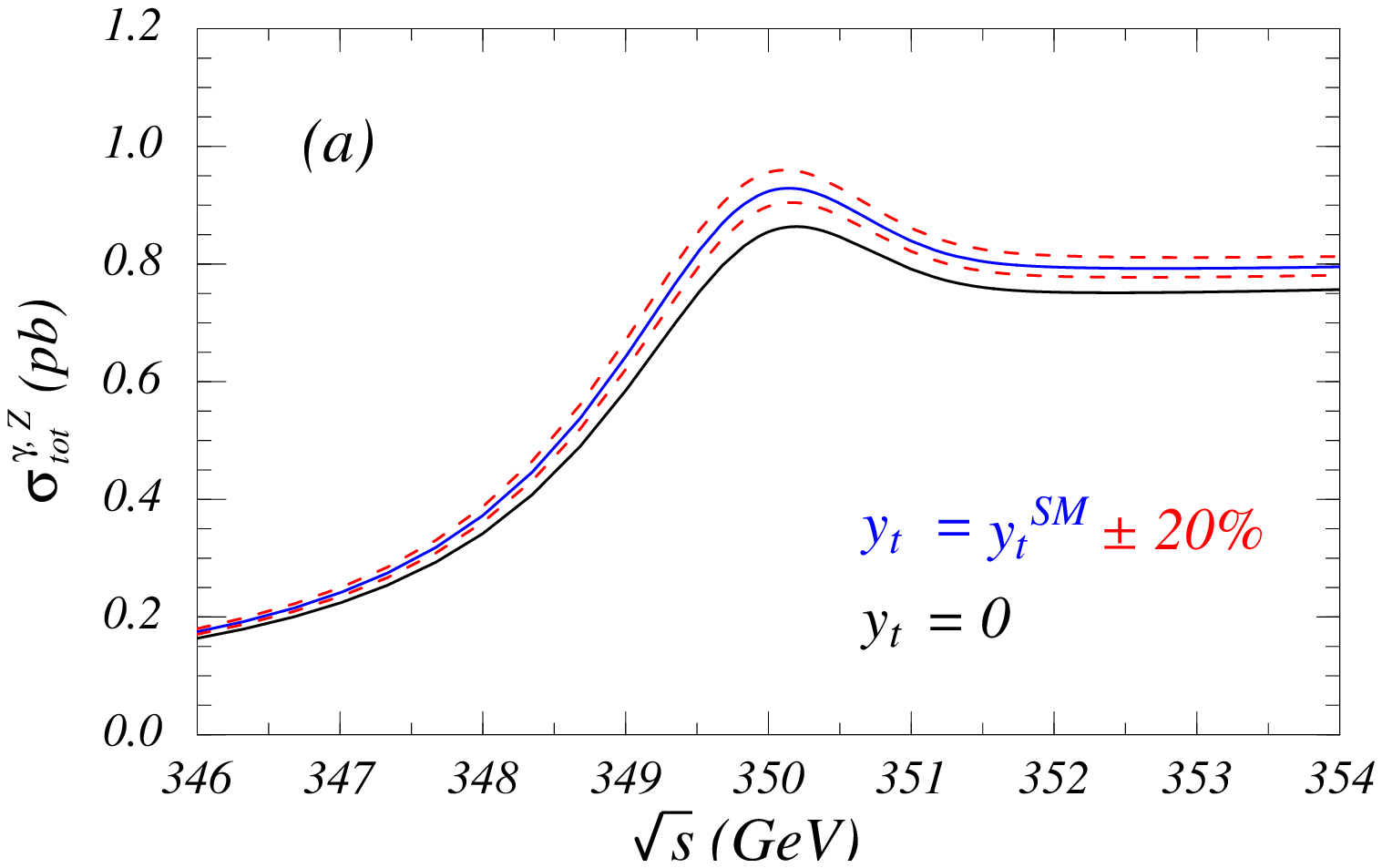}\hspace{0.2cm}
  \includegraphics[width=0.5\textwidth]{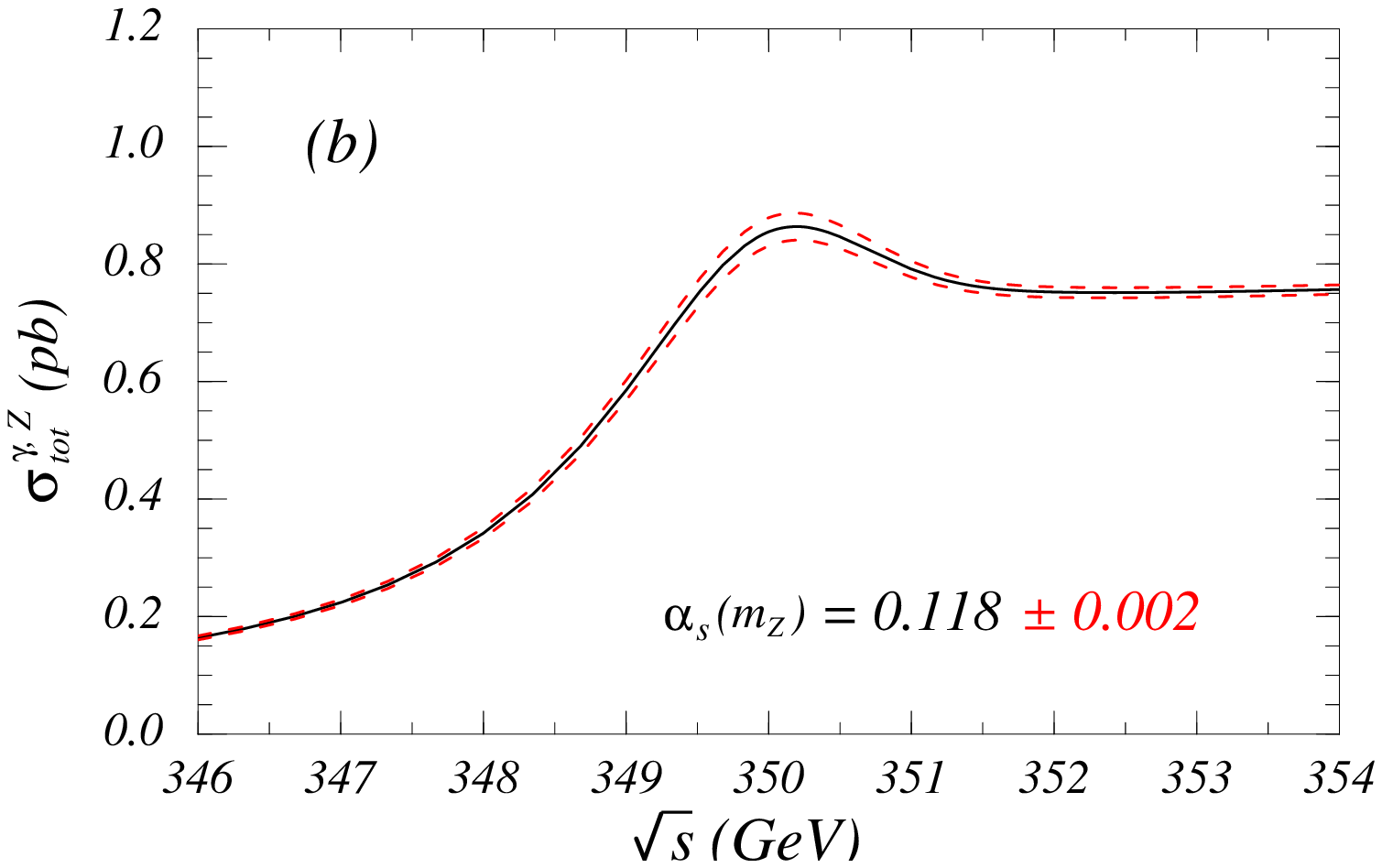}} \caption{Variation of the
  NNLL cross section for (a) the inclusion of a Standard Model (SM) Higgs boson
  and (b) the value of the strong coupling. The relative changes are shown by
  red dashed lines. For (a) the lower black solid line is
  the decoupling limit for the Higgs boson, and the upper blue solid line is for
  a SM Higgs with mass $m_H=115\,{\rm GeV}$.} \label{fig_params2}
\end{figure}
In Fig.~\ref{fig_params2} the dependence of the cross section on the top Yukawa
$y_t$ (for a Higgs mass $m_H=115\,{\rm GeV}$) and on $\alpha_s(m_Z)$ are
shown. It looks quite promising that a $\pm 20\%$ variation in $y_t$ gives a
larger change in the cross section than our estimate for the remaining
theoretical uncertainty. It should be kept in mind that since both $y_t$ and
$\alpha_s(m_Z)$ mainly effect the normalization, at some level these parameters
cannot be fixed independently using only the total cross section.

\section{Conclusion}

In this talk I have discussed predictions for the threshold $e^+e^-\to t\bar t$
cross section at NNLO and NNLL order as defined by the expansions in
Eq.~(\ref{RNNLO}) and Eq.~(\ref{RNNLL}). The NNLL predictions made in
Refs.~\cite{hmst,hmst2} sum large logarithms of the velocity using results for
the renormalization group improved Wilson coefficients from
Refs.~\cite{LMR,amis,amis3,hms}. One missing ingredient is the three loop
anomalous dimension for $c_1$, for which only partial results are
known. However, rough estimates indicate that this missing anomalous dimension
is unlikely to affect the cross section at more than the 2\%
level~\cite{hmst2}. The stability of predictions for the peak in the cross
section are very similar at NNLO and NNLL, so that measurements with $\delta m_t
< 200\,{\rm MeV}$ for short distance masses are feasible with either
expansion. The size of the NNLL normalization corrections and variation of the
NNLL cross section for various choices of the renormalization parameter are an
order of magnitude smaller than the results of earlier NNLO calculations.  A
conservative estimate of the remaining theoretical uncertainty in the total
cross section is $\pm 3\%$~\cite{hmst2}.  With such small uncertainty,
measurements of top parameters with uncertainties $\delta\alpha_s(m_Z)\sim
0.002$, $\delta\Gamma_t/\Gamma_t\sim 5\%$, and $\delta y_t/y_t\sim 20\%$ appear
feasible. However, realistic simulation studies should be done to see how these
numbers hold up once effects such as initial state radiation, beamstrahlung, and
the beam energy spread are taken into account.

\begin{theacknowledgments}

I would like to thank A.~Hoang, A.~Manohar, and T.~Teubner for their
collaboration on the results presented here.

\end{theacknowledgments}


\doingARLO[\bibliographystyle{aipproc}]
          {\ifthenelse{\equal{\AIPcitestyleselect}{num}}
             {\bibliographystyle{arlonum}}
             {\bibliographystyle{arlobib}}
          }
\bibliography{hf9-proc}

\end{document}